\documentclass[aps,prd,nofootinbib]{revtex4}
\usepackage[latin1]{inputenc}
\usepackage{float}
\usepackage{amsmath}
\usepackage{amsfonts}
\usepackage{graphicx}
\usepackage{hyperref}
\usepackage{wasysym}

\def\be{\begin{equation}}
\def\ee{\end{equation}}
\def\bea{\begin{eqnarray}}
\def\eea{\end{eqnarray}}

\def\be{\begin{equation}}
\def\ee{\end{equation}}
\def\bea{\begin{eqnarray}}
\def\eea{\end{eqnarray}}

\def\G{{\cal G}}

\def\be{\begin{equation}}
\def\ee{\end{equation}}
\def\bea{\begin{eqnarray}}
\def\eea{\end{eqnarray}}

\def\5{\overline 5}

\newcommand{\swa}{Schwarzschild}
\begin{document}

\title{Construction of cosmologically viable $f(\G)$ gravity models}

\author{Antonio De Felice}
\affiliation{Theoretical and Mathematical Physics Group, Centre for
  Particle Physics and Phenomenology, Louvain University, 2 Chemin du
  Cyclotron, 1348 Louvain-la-Neuve (Belgium)}

\author{Shinji Tsujikawa} \affiliation{Department of Physics, Faculty
  of Science, Tokyo University of Science, 1-3, Kagurazaka,
  Shinjuku-ku, Tokyo 162-8601, Japan} \email{shinji@rs.kagu.tus.ac.jp}
\date{\today}

\vskip 1pc

\begin{abstract}

  We derive conditions under which $f(\G)$ gravity models, whose
  Lagrangian densities $f$ are written in terms of a Gauss-Bonnet term
  $\G$, are cosmologically viable.  The most crucial condition to be
  satisfied is ${\rm d}^2 f/{\rm d} \G^2>0$, which is required to
  ensure the stability of a late-time de-Sitter solution as well as
  the existence of standard radiation/matter dominated epochs.  We
  present a number of explicit $f(\G)$ models in which a cosmic
  acceleration is followed by the matter era.  We find that the
  equation of state of dark energy can cross the phantom divide before
  reaching the present Universe.  The viable models have asymptotic
  behavior ${\rm d}^2 f/{\rm d} \G^2 \to +0$ for $|\G| \to \infty$, in
  which case a rapid oscillation of perturbations occurs unless such
  an oscillating degree of freedom is suppressed relative to a
  homogeneous mode in the early universe.  We also introduce an
  iterative method to avoid numerical instabilities associated with a
  large mass of the oscillating mode.

\end{abstract}


\maketitle
\vskip 1pc

%%%%%%%%%%%%%%%%%%%%%%%%%%
\section{Introduction}
%%%%%%%%%%%%%%%%%%%%%%%%%%

The late-time cosmic acceleration can, in principle, originate from a
modification of gravity rather than an exotic source of matter with a
negative pressure. Over the past five years, a lot of works on
modified gravity have been done to identify the origin of dark energy
(DE) \cite{review}.  The attractive point in modified gravity models
is that they are generally more strongly constrained from cosmological
observations and local gravity experiments than the models based on
the exotic source of matter.

Presumably the simplest extension to Einstein gravity is the so-called
$f(R)$ gravity in which $f$ is an arbitrary function of the Ricci
scalar $R$ \cite{Capo}.  Even in this simple case it is not generally
easy to construct viable $f(R)$ models that are consistent with
cosmological and local gravity constraints.  The main reason for this
is that $f(R)$ gravity gives rise to a strong coupling between DE and
a non-relativistic matter in the Einstein frame \cite{APT}.  The
models need to be carefully designed so that a scalar degree of
freedom (``scalaron'' \cite{Star80}) is nearly frozen to suppress an
effective coupling between the scalar field and matter.

The conditions for the cosmological viability of $f(R)$ models have
been derived in Ref.~\cite{AGPT}.  Among those conditions the
requirement, ${\rm d}^2 f/{\rm d} R^2>0$, is particularly important to
give rise to a saddle matter era followed by a late-time cosmic
acceleration. This is also required for the stability of cosmological
perturbations \cite{per} as well as for the consistency with local
gravity experiments \cite{lgc}.  The cosmologically viable $f(R)$
models need to be close to the $\Lambda$-Cold Dark Matter
($\Lambda$CDM) model in the deep matter era, but the deviation from it
becomes important around the late stage of the matter era.  Several
examples of such viable models were presented in Refs.~\cite{Li,AT08}.

Hu and Sawicki \cite{Hu07} and Starobinksy \cite{Star07} proposed
$f(R)$ models that are consistent with local gravity constraints as
well as cosmological constraints (see also Refs.~\cite{otherref}).  In
these models it is possible to find an appreciable deviation from the
$\Lambda$CDM cosmology.  This leaves a number of interesting
observational signatures such as the peculiar evolution of the DE
equation of state \cite{Tsuji07,DMW}, the modification of the matter
power spectrum \cite{Star07,Tsuji07,SPH,Oya} and the change of the
convergence spectrum of weak lensing \cite{WL,Schmidt}.  This is a
welcome feature to distinguish $f(R)$ models from the $\Lambda$CDM
cosmology in future observations.

There are other modified gravity DE models that are the
generalizations of $f(R)$ gravity.  For example Carroll {\it et al.}
\cite{Carroll04} proposed theories with the Lagrangian density $f(R,
P, Q)$, where $f$ is an arbitrary function of $R$, $P \equiv R_{\mu
  \nu}R^{\mu \nu}$ and $Q \equiv R_{\mu \nu \rho \sigma} R^{\mu \nu
  \rho \sigma}$ (here $R_{\mu \nu}$ and $R_{\mu \nu \rho \sigma}$ are
the Ricci tensor and the Riemann tensor, respectively).  These
theories are plagued by the appearance of spurious spin-2 ghosts
unless a Gauss-Bonnet (GB) combination, i.e., $f=f(R, Q-4P)$, is
chosen \cite{ghost1,ghost2,Cal05,Cal}.  Even in this case the graviton
itself may still become a ghost in the
Friedmann-Lemaitre-Robertson-Walker (FLRW) spacetime, unless some
no-ghost conditions are verified on the background \cite{Cal}.

The GB curvature invariant Lagrangian, $\G \equiv R^2-4R_{\mu
  \nu}R^{\mu \nu}+R_{\mu \nu \rho \sigma} R^{\mu \nu \rho \sigma}$, is
a total derivative in the 4-dimensional FLRW background.  In order to
give rise to some contribution of the GB term to the Friedmann
equation, we require that (i) the GB term couples to a scalar field
$\phi$, i.e., $F(\phi)\G$, or (ii) the Lagrangian density $f$ is a
function of $\G$, i.e., $f(\G)$.  The GB coupling in the case (i)
appears in low-energy string effective action \cite{string} and
cosmological solutions in such theory have been studied in great
details \cite{GBpapers}.  It was shown by several authors that a
late-time cosmic acceleration following a (scaling) matter era occurs
for an exponential coupling $F(\phi) \propto e^{\lambda \phi}$ in the
presence of a scalar-field potential
\cite{NO05,Koi,TS,Sanyal,Neupane}.  Amendola {\it et al.} \cite{Amen}
studied local gravity constraints in such models and showed that the
energy contribution coming from the GB term needs to be strongly
suppressed for the compatibility with solar-system experiments. Thus,
in the case (i), it is generally difficult to satisfy local gravity
constraints if the GB term is responsible for DE.

In the context of $f(\G)$ gravity there exists a de-Sitter point that
can be used for cosmic acceleration \cite{O05} (see also Ref.~\cite{Nojiri07}).  
It was shown in Ref.~\cite{DeFe} that the model with inverse powers of linear
combinations of quadratic curvature invariants, i.e., $f(\G)=\G^{n}$
with $n<0$, is not cosmologically viable because of the presence of
separatrices between radiation and DE dominations.  In
Ref.~\cite{Davis} it was found that the model $f(\G)=\G^n$ with $n>0$
can be consistent with solar-system tests for $n \lesssim 0.074$ if
the GB term is responsible for DE.  Li {\it et al.} \cite{LBM} showed
that it is difficult to reproduce standard expansion history of the
Universe unless $f(\G)$ is close to cosmological constant.

In this paper we present a number of explicit models of $f(\G)$
gravity that are cosmologically viable. These models mimic the
$\Lambda$CDM cosmology in the deep matter era, but the deviation from
it becomes important at late times on cosmological scales.  This
situation is similar to the viable $f(R)$ models proposed by Hu and
Sawicki \cite{Hu07} and Starobinsky \cite{Star07}.  In $f(\G)$
gravity, however, the GB term changes its sign during the transition
from the matter era to the accelerated epoch.  We need to take into
account this property when we construct viable $f(\G)$ models.  For
example, the $f(\G)$ model that replaces $R$ in the model of Hu and
Sawicki or Starobinsky for $\G$ is not cosmologically viable. 

There is another difference between $f(R)$ and $f(\G)$ theories.  The
Ricci scalar $R$ vanishes for the vacuum \swa~solution, whereas the GB
term has a non-vanishing value much larger than $H_0^4$ around the
compact objects \cite{Davis} ($H_0$ is the present Hubble parameter).
In the presence of matter with a density $\rho_m$ the Ricci scalar $R$
is roughly of the order of $\rho_m$ so that one has $R/H_0^2 \sim
\rho_m/\rho_c$, where $\rho_c$ is the present cosmological density.
The viable $f(R)$ models \cite{Hu07,Star07} are designed to have the
suppression term $(R/H_0^2)^{-n}$~($n>0$) in addition to the
$\Lambda$CDM Lagrangian in the region of high density ($\rho_m \gg
\rho_c$).  In the $f(\G)$ gravity this sort of suppression occurs even
in the vacuum background because of the condition $\G\gg H_0^4$.
Hence the $f(\G)$ models might be less constrained by local gravity
constraints relative to the $f(R)$ models.  For the same reason, one
could expect that for the interior star solutions these modifications
of gravity could remain small corrections, provided that singularities
of the kind $f_{,\G\G} \equiv {\rm d}^2 f/{\rm d} \G^2=0,\infty$ for
finite values of $\G$ are not encountered.  The models discussed in
this paper have exactly this feature.  Note, however, that a detailed
study of these issues is needed in order to to further constrain these
modifications, which we leave for future work.

We will show that the stability condition for a late-time de-Sitter
point is given by $f_{,\G \G}>0$.  This can be also derived by
considering the stability of radiation and matter points.  Note that
the same condition has been derived in Ref.~\cite{LBM} by studying the
evolution of cosmological perturbations.  Since Li {\it et al.}
\cite{LBM} used the metric signature $(+,-,-,-)$ instead of
$(-,+,+,+)$ that we adopt throughout this paper, their stability
condition $f_{,\G \G}<0$ corresponds to our stability condition
$f_{,\G \G}>0$.

For the viable $f(\G)$ models we propose in this paper, the second
derivative $f_{,\G \G}$ approaches $+0$ as $|\G|$ gets larger. In
this case the perturbations in the Hubble parameter $H$ have a large
mass squared proportional to $1/(H^4f_{,\G \G})$ during radiation and
matter eras.  This can give rise to rapid oscillations of the Hubble
parameter and matter density perturbations, as they happen for viable
models in $f(R)$ gravity \cite{Star07,Tsuji07}.  In order to avoid
this, the oscillating mode needs to be suppressed relative to the
homogeneous mode in the early Universe.  The numerical instability we
typically face in radiation and matter epochs is associated with the
appearance of this oscillating mode.  We shall introduce an iterative
method to avoid this numerical instability in the high-redshift
regime.

%%%%%%%%%%%%%%%%%%%%%%%%%%%%%%%%
\section{Cosmologically viable $f(\G)$ dark energy models}
\label{secmodel}
%%%%%%%%%%%%%%%%%%%%%%%%%%%%%%%%

We start with the following action
\begin{eqnarray}
\label{action}
S =  \int {\rm d}^4 x\sqrt{-g} 
\left[ \tfrac12\, R +f(\G) \right]
+S_m (g_{\mu \nu}, \Psi_m)\,,
\end{eqnarray}
where $R$ is a Ricci scalar, and $f$ is a general differentiable
function of $\G$, $S_m$ is a matter action that depends on a spacetime
metric $g_{\mu \nu}$ and matter fields $\Psi_m$.  We choose units such
that $\kappa^2 \equiv 8\pi G_N=1$, where $G_N$ is a bare gravitational
constant.  It should be pointed out that this action, at the classical
level, can be rewritten as an auxiliary scalar field coupled to the
$\G$ term, as shown in Ref.~\cite{Cal}, following a trick used for the
$f(R)$ theory.  However there is no conformal transformation
separating $\G$ from such a field, unlike the $f(R)$ theory in which
the conformal transformation leads to an Einstein frame action with a
canonical scalar field coupled to matter.  An important quantity in
$f(\G)$ gravity is $1/f_{,\G\G}$, which plays the role of an effective
mass for the theory (as we shall see later).

The variation of the action (\ref{action}) with respect to $g_{\mu \nu}$
leads to the following equation
\begin{eqnarray}
\label{geeq}
G_{\mu \nu}+8 \left[ R_{\mu \rho \nu \sigma} +R_{\rho \nu} g_{\sigma \mu}
-R_{\rho \sigma} g_{\nu \mu} -R_{\mu \nu} g_{\sigma \rho}+
R_{\mu \sigma} g_{\nu \rho}+\tfrac{R}{2} (g_{\mu \nu} g_{\sigma \rho}
-g_{\mu \sigma} g_{\nu \rho}) \right] \nabla^{\rho} \nabla^{\sigma} f_{,\G}
+(\G f_{,\G}-f) g_{\mu \nu}=T_{\mu \nu}\,,
\end{eqnarray}
where $G_{\mu \nu}=R_{\mu \nu}-(1/2)R g_{\mu \nu}$ is the Einstein-tensor.  
For the energy momentum tensor $T_{\mu \nu}$ of a matter fluid
we take into account the contributions of non-relativistic matter 
(energy density $\rho_m$) and radiation (energy density $\rho_{\rm rad}$).
In a flat FLRW background with the metric
${\rm d}s^2=-{\rm d}t^2+a(t)^2 {\rm d}{\bf x}^2$, 
the 00 component of Eq.~(\ref{geeq}) gives
\begin{eqnarray}
\label{be1}
3H^2=\G f_{,\G}-f-24H^3 \dot{f_{,\G}}+
\rho_m+\rho_{{\rm rad}}\,,
\end{eqnarray}
where $H \equiv \dot{a}/a$, $f_{,\G} 
\equiv {\rm d}f/{\rm d} \G$, and a dot represents a derivative 
with respect to cosmic time $t$.
The GB term is given by 
\begin{eqnarray}
\label{GBterm}
\G=24H^2 (H^2+\dot{H})\,.
\end{eqnarray}
The energy densities $\rho_m$ and $\rho_{\rm rad}$
satisfy the continuity equations $\dot{\rho}_m+3H \rho_m=0$
and $\dot{\rho}_{\rm rad}+4H \rho_{\rm rad}=0$, respectively.

\subsection{Stability of de Sitter point}

Let us first discuss the stability around a de Sitter point 
present in $f(\G)$ gravity
by neglecting the contribution of pressure-less matter 
and radiation.
The Hubble parameter, $H=H_1$, 
at the de Sitter point satisfies
\begin{eqnarray}
\label{deSi}
3H_1^2=\G_1 f_{,\G}(\G_1)-f(\G_1)\,, 
\end{eqnarray}
where $\G_1=24H_1^4$.
Note that we used the relations $\dot{H}_1=0$ and $\dot{\G}_1=0$.
Considering a linear perturbation $\delta H_1$ about the de Sitter point,
Eq.~(\ref{be1}) gives
\begin{eqnarray}
\label{deltaH1}
\delta H_1=4H_1^2 f_{,\G \G}(H_1) \left[H_1 \delta \G (H_1)
-\delta \dot{\G} (H_1) \right]\,.
\end{eqnarray}
Substituting the relations $\delta \G (H_1)=24 (4H_1^3 \delta H_1
+H_1^2 \dot{\delta{H}_1})$ and 
$\delta \dot{\G} (H_1)=24H_1^2 (\ddot{\delta{H}_1}+4H_1 \dot{\delta H_1})$
into Eq.~(\ref{deltaH1}), we obtain
\begin{eqnarray}
\label{desitter}
\ddot{\delta{H}}_1+3H_1 \dot{\delta{H}}_1 +\left[ \frac{1}{96H_1^6 
f_{,\G \G}(H_1)}-4\right] H_1^2 \delta H_1=0\,.
\end{eqnarray}
This shows that the effective mass squared is $[(96H_1^6 f_{,\G
  \G}(H_1))^{-1}-4]H_1^2$.  The solution to Eq.~(\ref{desitter}) is
given by
\begin{eqnarray}
\delta H_1=c_1 e^{\lambda_+ t}+c_2 e^{\lambda_- t}\,,
\qquad
\lambda_\pm=\frac{3H_1}{2} \left[ -1 \pm
\sqrt{1-\frac49 \left( \frac{1}{96H_1^6 
f_{,\G \G}(H_1)}-4 \right)} \right]\,,
\end{eqnarray}
where $c_1$ and $c_2$ are integration constants.
This shows that the de Sitter point is stable under the condition 
\begin{eqnarray}
\label{decon}
0<H_1^6 f_{,\G \G} (H_1) <1/384\,,
\end{eqnarray}
which requires that $f_{,\G \G} (H_1)>0$. Especially when $0<H_1^6 f_{,\G \G} (H_1) <1/600$, the de Sitter 
point corresponds to a stable spiral (damping with oscillations).

\subsection{Stabilities of radiation and matter points}

We shall also study the stability of matter and radiation fixed points 
by using a similar method discussed above. 
Let us investigate the case in which the evolution of 
the scale factor is given by $a \propto t^p$ ($p$:\,constant) 
with the dominance of a fluid characterized by an energy 
density $\rho_M$ (either $\rho_m$ or $\rho_{\rm rad})$. Then we have 
\be
\label{background}
3H^2=\G f_{,\G}-f-24H^3 f_{,\G \G} \dot{\G}+\rho_M\,.
\ee
We consider the first-order perturbations 
$\delta_H$ and $\delta_M$ as follows:
\be
H=H^{(b)} (1+\delta_H)\,,\quad
\rho_M=\rho_M^{(b)} (1+\delta_M)\,.
\ee
Here the subscript ``(b)'' represents background values, 
but in the following we omit it for simplicity.
Taking the homogeneous perturbations of Eq.~(\ref{background}) and 
using the approximate relation $3H^2 \simeq \rho_M$, we find
\begin{align}
\label{geper}
\ddot{\delta_H}&+\left[ 3-\frac{6}{p}+\frac{96H^4 f_{,\G \G \G}}
{f_{,\G \G}}
\frac{1-p}{p^2} \right] H \dot{\delta_H}
+\left[ \frac{21(1-p)}{p^2}+\frac{1}{96H^6 f_{,\G \G}}-4+
\frac{96H^4 f_{,\G \G \G}}{f_{,\G \G}} \left( 4-\frac{3}{p} \right)
\frac{1-p}{p^2} \right] H^2 \delta_H \nonumber \\
&=\frac{\delta_M}{192 H^6 f_{,\G \G}}H^2.
\end{align}
In the limit $p \to \infty$ without matter perturbations,
Eq.~(\ref{desitter}) is recovered.

The solution to Eq.~(\ref{geper}) is described by the sum of 
the matter-induced mode $\delta_H^{({\rm ind})}$ and the oscillating mode 
$\delta_H^{({\rm osc})}$ \cite{Star07,Tsuji07}:
\be
\delta_H=\delta_H^{({\rm ind})}+\delta_H^{({\rm osc})}\,.
\ee
The matter-induced mode corresponds to a special solution to 
Eq.~(\ref{geper}) induced by the matter perturbation $\delta_M$.
The oscillating mode is the solution of the equation 
with $\delta_M=0$ in Eq.~(\ref{geper}).
For the $f(\G)$ models whose deviation from the $\Lambda$CDM model is
small during radiation and matter eras, $f_{,\G \G}$ is close to 0.
In this case the mass squared
\be
\label{masssqu}
M^2 \equiv \frac{1}{96 H^4 f_{,\G \G}}\,,
\ee
is the dominant contribution in front of the term $\delta_H$
in Eq.~(\ref{geper}).
In order to avoid a violent instability of perturbations we require 
that $M^2>0$, giving the condition $f_{,\G \G}>0$.
In this case the perturbation $\delta_H^{({\rm osc})}$ oscillates
with a frequency of the order of $M$.

During radiation and matter dominated epochs the GB term
evolves as $\G=-24 H^4$ and $\G=-12H^4$, respectively.
Since $\G=24H^2 (\ddot{a}/a)$ from Eq.~(\ref{GBterm}), 
the GB term changes its sign at the onset of the late-time acceleration.
Hence the condition $f_{,\G \G}>0$ needs to be satisfied
in the region $\G \le \G_1$.
Since the term $24H^3 f_{,\G \G} \dot{\G}$ on the r.h.s. of Eq.~(\ref{background})
is of the order of $H^8 f_{,\G \G}$, this is suppressed relative to 
$3H^2$ under the condition $H^6 f_{,\G \G} \ll 1$.
In order for this condition to hold in the radiation and matter eras, 
we require that the term $f_{,\G \G }$ approaches $+0$ with the
increase of $|\G|$.

\subsection{Viable $f(\G)$ models}

If we consider a spherically symmetric body (mass $M_{\odot}$ and 
radius $r_{\odot}$) with a homogeneous density, 
it was shown in Ref.~\cite{Davis} that the GB term inside and outside
the body is given by $\G=-48(G_N M_{\odot})^2/r_{\odot}^6$ 
and $\G=48(G_N M_{\odot})^2/r^6$, respectively
($r$ is a distance from the center of symmetry).
In the vicinity of the Sun or the Earth,  $|\G|$ is much larger than  
the present cosmological GB term, $\G_0$.
As we move from the interior to the exterior of the star, 
the GB term crosses 0 from negative to positive.
This means that $f(\G)$ and its derivatives with respect to $\G$
need to be regular for both negative and positive values of $\G$ 
whose amplitudes are much larger than $\G_0$.

{}From the above discussions the viable models need to satisfy the
following conditions: 
\begin{itemize}
\item (i) $f(\G)$ and its derivatives $f_{,\G}$, $f_{, \G\G}$,...
are regular. 
\item (ii) $f_{,\G \G}>0$ for $\G \le \G_1$ and $f_{,\G \G}$ 
approaches $+0$ in the limit $|\G| \to \infty$.
\item (iii) $0<H_1^6 f_{,\G \G} (H_1)<1/384$ at the de Sitter point.
\end{itemize}

A number of examples for the viable forms of $f_{,\G \G}$ are 
\be
\label{fGG}
{\rm (a)}~f_{,\G \G}=\frac{\lambda}{{\G_*}^{3/2}\left[ 1+(\G^2/\G_*^2)^{n} \right]},\quad
{\rm (b)}~f_{,\G \G}=\frac{2\lambda}{{\G_*}^{3/2}\left( 1+\G^2/\G_*^2 \right)^n}\,,
\quad
{\rm (c)}~f_{,\G \G}=\frac{\lambda}{{\G_*}^{3/2}}\left[1-\tanh^2(\G/\G_*)\right]\,,
\ee
where $\lambda$, $n$ and $\G_*$ are positive constants.
These satisfy the condition $f_{,\G \G}>0$ for all values of $\G$.
In the following let us study the case (a) with $n=1$, the case (b) with $n=2$, 
and the case (c). 
Integrating $f_{, \G \G}$ with respect to $\G$ twice, we obtain 
the following models
\bea
\label{modela}
& &{\rm (A)}~f (\G)=\lambda \frac{\G}{\sqrt{\G_*}}\,{\rm arctan} \left( \frac{\G}{\G_*} \right)
-\frac{\lambda}{2} \sqrt{G_*}\,{\rm ln} \left(1+\frac{\G^2}{\G_*^2} \right)
-\alpha \lambda \sqrt{\G_*}
\,,\\
& &{\rm (B)}~f (\G)=\lambda \frac{\G}{\sqrt{\G_*}}\,{\rm arctan} \left( \frac{\G}{\G_*} \right)
-\alpha \lambda \sqrt{\G_*}
\,,\\
\label{modelc}
& &{\rm (C)}~f (\G)=\lambda \sqrt{\G_*}\,{\rm ln} \left[ \cosh \left( \frac{\G}{\G_*} \right) \right]
-\alpha \lambda \sqrt{\G_*}\,,
\label{model}
\eea
where $\alpha$ is a constant.
We have dropped the terms proportional to $\G$, since they do not give rise to 
any contribution to the evolution equations.
If we demand the condition $f(\G=0)=0$ then we have $\alpha=0$.
These models are also consistent with the regularity condition (i).
Note that the model $f(\G)=-\lambda \G_c^{1/2} [1- (1+\G^2/\G_c^2)^{-n}]$,
which is a generalization of the viable $f(R)$ model proposed by Starobinsky \cite{Star07},  
is not compatible with the condition $f_{,\G \G}>0$ for $\G<0$.

In the models (A), (B), (C) the term $\G f_{,\G}-f$ on the r.h.s. of Eq.~(\ref{be1})
in the regime $|\G| \gg \G_*$ can be estimated by 
(A) $\G f_{,\G}-f \simeq (\lambda/2) \sqrt{\G_*}\, \left[{\rm ln}(\G^2/\G_*^2)+2\alpha \right]$,
(B) $\G f_{,\G}-f \simeq \lambda \sqrt{\G_*} (1+\alpha)$,
(C)  $\G f_{,\G}-f \simeq \lambda \sqrt{\G_*}({\rm ln}\,2+\alpha)$, respectively.
As long as $G_*$ is of the order of $H_0^4$ (where 
$H_0$ is the present Hubble parameter) the term $\G f_{,\G}-f$ is 
subdominant relative to the term $3H^2$ during radiation and matter 
eras (for $\alpha \lambda$ of the order of unity).
Moreover we have $H^6 f_{,\G \G} \ll 1$ for $|\G| \gg \G_*$ in the above
three models, which means that the condition $|24H^3 f_{,\G \G} \dot{\G}| \ll 3H^2$
is satisfied in this regime.
The contribution coming from the $f(\G)$ term becomes important when 
$H$ decreases to the order of $H_*$.

Let us discuss the condition (iii) for the model (A).
At the de Sitter point this model satisfies the following relation
\be
\label{lamvalue}
\lambda=\frac{6y_1^2}{2\alpha+{\rm ln} (1+24^{2} y_1^8)}\,,\quad
{\rm where}\quad  y_1=\frac{H_1}{\G_*^{1/4}}\,.
\ee
When $\alpha=0$ the r.h.s. has a minimum value $\lambda_{\rm min}=0.828$ at 
$y_1=0.736$. Hence two de Sitter points exist for 
$\lambda>0.828$.
The stability condition for the de Sitter solution corresponds to 
\be
\label{stacon}
\lambda<\frac{1+24^2 y_1^8}{384 y_1^6}\,,
\ee
which gives $y_1>0.736$ by using Eq.~(\ref{lamvalue}).
Hence one of the de Sitter points that exists in the region 
$y_1>0.736$ is stable.

In the case where $\alpha=0$ we have numerically found instabilities 
of cosmological solutions around the region $\G=0$ (which occurs
during the transition from the matter era to the accelerated era).
One can estimate the stability of solutions by setting $p=1$
in Eq.~(\ref{geper}). This gives the stability condition
\be
0<H_2^6 f_{,\G \G} (H_2)<1/384\,,
\ee
where $H_2$ is the Hubble parameter at $\G=0$.
For the model (A) this translates into 
\be
\label{lamsec}
\lambda<\frac{1}{384 y_2^6}\,,
\ee
where $y_2=H_2/\G_*^{1/4}$.  Here $y_2$ is slightly larger than $y_1$.
Since $y_1>0.736$ for $\alpha=0$, the condition (\ref{lamsec})
requires that $\lambda \ll 1$.  However this is not compatible with
the condition $\lambda>0.828$ for the existence of de Sitter
solutions.  As we will see in the next section, it is generally
difficult to have a natural transition around $\G=0$ for $\alpha=0$.
This anticipates that the cosmological constant term may be 
needed in general for the cosmological viability of the $f(\G)$ models.
If this is the case, the need of such a constant term
makes the $f(\G)$ models less attractive from a theoretical point of view.

If $\alpha \neq 0$ it is possible to make $\lambda$ much smaller than 1
provided that $\alpha \gg 1$ [see Eq.~(\ref{lamvalue})].
When $\alpha=10^2$ and $y_1=0.1$, for example, we have 
$\lambda=3.0 \times 10^{-4}$ and hence the condition (\ref{lamsec})
is satisfied even for $y_2=1$.
In the next section we shall show that viable cosmological trajectories
can be realized for $\alpha$ larger than the order of unity.

The stabilities of cosmological solutions for the models (B) and (C)
are similar to those for the model (A) discussed above.
The difference is that the models (B) and (C) have larger powers of 
$\G^2/\G_*^2$ in $f_{,\G \G}$ compared to the model (A).
Since the mass $M^2$ grows rapidly toward the past
in such models, it is more difficult to start solving the 
equations numerically from the high-redshift regime
unless we use an iterative method we discuss later.

\subsection{Oscillating modes in the early Universe}

At the end of this section we discuss the evolution of the homogeneous
perturbation $\delta_H$ during radiation and matter eras for the
viable $f(\G)$ models presented above. For the matter-induced mode,
the mass term $M^2 \delta_H$ balances with the source term on the
r.h.s. of Eq.~(\ref{geper}), giving
\be
\delta_H^{({\rm ind})} \simeq \delta_M/2\,.
\ee
Hence the matter-induced mode grows in proportion to $\delta_M$.  In
the matter era $\delta_H^{({\rm ind})} \propto t^{2/3}$ in the regime
where the model is close to $\Lambda$CDM model \cite{Tsuji07}.

Let us consider the model (B) for the evolution of the oscillating
mode in the regime $|\G| \gg \G_*$.  In this case we have $96H^4
(f_{,\G \G \G}/f_{,\G \G})(1-p)/p^2 \simeq 16/p$ and $M^2 \simeq \mu^2
t^{-12}$, where $\mu$ is a constant.  Hence the oscillating mode
satisfies
\be
\ddot{\delta_H}^{({\rm osc})} +\frac{ 3p+10}{t}\,
\dot{\delta_H}^{({\rm osc})}+
\frac{\mu^2}{t^{12}}\, 
\delta_H^{({\rm osc})} \simeq 0\,.
\ee
The solution to this equation can be written as the combination of
Bessel differential functions:
\be
\label{pert0}
\delta_H^{({\rm osc})}=\left(\frac{t_{\rm i}}{t}\right)^{\!1+5/p}
\left[ \frac{\delta_{H,t_{\rm i}}^{\{1\}}}{J_{\theta_1}(z_{\rm i})}\,J_{\theta_1}(z)+
\frac{\delta_{H,t_{\rm i}}^{\{2\}}}{J_{-\theta_1}(z_{\rm i})}\,J_{-\theta_1}(z)\right]\, ,
\ee
where $\theta_1=(5+p)/(5p)$, $z=\mu/(5t^{5})$, $z_{\rm i}=z(t_{\rm
  i})$, and $t_{\rm i}$ is the initial time at which two modes in the
square bracket of Eq.~(\ref{pert0}) have amplitudes $\delta_{H,t_{\rm
    i}}^{\{1\}}$ and $\delta_{H,t_{\rm i}}^{\{2\}}$.

As $t \to 0$, the solution reduces to 
\bea
\delta_H^{({\rm osc})}&\simeq& A\,t^{-(10-3p)/(2p)}
\left\{C_1\cos\!\left[\frac{\pi}{4}\,(1+2\theta_1)-\frac{\mu\,
      t^{-5}}{5}\right]+C_2\cos\!\left[\frac{\pi}{4}\,(1-2\theta_1)-\frac{\mu\,
      t^{-5}}{5}\right]\right\}\,,  
\eea 
where $A=\mu^{-1/4}\,\sqrt{10/\pi}$, and $C_1$ and $C_2$
are constants.
During the matter era ($p=2/3$), the amplitude of $\delta_H^{({\rm osc})}$ 
is proportional to $t^{-6}$ so that the oscillating mode tends to be negligible 
relative to the matter-induced mode with time. 
However, as we go back to the past, the amplitude 
grows with larger frequency of oscillations.
Since one has ${\rm Amp} \bigl[\delta_H^{({\rm osc})}\bigr]
\propto t^{-17/2}$ for $p=1/2$,
this property also persists during the radiation-dominated epoch.
As we see in Sec.~\ref{dynamics} the large mass term $M$
tends to lead numerically instabilities associated with violent oscillations
of $\delta_H$, unless the oscillating mode is strongly suppressed
relative to the matter-induced mode.
For the model (A) the growth of the mass squared is not so strong
($M^2 \propto t^{-4}$) relative to the model (B), but still 
it is difficult to solve equations numerically from the high-redshift regime.
This property is even severe for the model (C) because 
of the rapid increase of the mass squared: $M^2 \propto t^4 \exp(c/t^4)$
($c$ is a positive constant).

%%%%%%%%%%%%%%%%%%%%%%%%%%%%%%%%
\section{Cosmological dynamics}
\label{dynamics}
%%%%%%%%%%%%%%%%%%%%%%%%%%%%%%%%

In this section we discuss cosmological dynamics for the viable models
presented in the previous section.
In subsection \ref{subA} we first integrate the background equations 
for the model (A) directly in the low-redshift regime.
It is more difficult to solve the equations numerically as we start integrating
from higher redshifts.  This is related to the appearance of the oscillating mode
with a large frequency. In subsection \ref{subB} we shall propose 
an iterative method to get approximate solutions in such a situation.

\subsection{Low-redshift cosmological solutions}
\label{subA}

In order to discuss cosmological solutions in the low-redshift regime, 
it is convenient to introduce the following dimensionless quantities
\bea
x \equiv \frac{\dot{H}}{H^2}\,,\quad
y \equiv \frac{H}{H_*}\,, \quad
\Omega_m \equiv \frac{\rho_m}{3H^2}\,,\quad
\Omega_{\rm rad} \equiv \frac{\rho_{\rm rad}}{3H^2}\,,
\eea 
where $H_*=G_*^{1/4}$.
We then obtain the following equations of motion
\bea
\label{beeq1}
& &x'=-4x^2-4x+\frac{1}{24^2H^6 f_{,\G \G}}
\left[ \frac{\G f_{,\G}-f}{H^2}
-3(1-\Omega_m-\Omega_{\rm rad}) \right]\,, \\
& &y'=xy\,,\\
& &\Omega_m'=-(3+2x)\Omega_m\,,\\
\label{beeq4}
& &\Omega_{\rm rad}'=-(4+2x)\Omega_{\rm rad}\,,
\eea 
where a prime represents a derivative with respect to $N=\ln(a)$.
The quantities $H^6 f_{,\G \G}$ and $(\G f_{,\G}-f)/H^2$ can be 
expressed by $x$ and $y$ once the model is specified.
The energy fraction of DE is given by 
$\Omega_{\rm DE} =1-\Omega_m-\Omega_{\rm rad}$.
We also define the effective equation of state
\bea
w_{\rm eff} \equiv -1-\frac{2\dot{H}}{3H^2}
=-1-\frac23 x\,,
\eea 
which changes from $0$ to $-1$ from the matter era
to the final de Sitter epoch for viable $f(\G)$ models.

\begin{figure}[ht]
 \begin{centering}\includegraphics[width=8.0truecm]{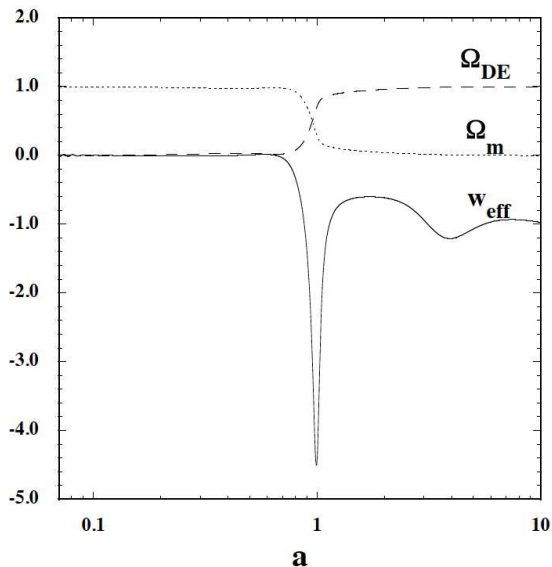} 
 \par\end{centering}
\caption{The evolution of $w_{\rm eff}$, $\Omega_{\rm DE}$ 
and $\Omega_m$ versus the scale factor $a$ for the model (A)
with parameters $\alpha=10.0$ and $\lambda=7.5 \times 10^{-2}$.
The initial conditions are chosen to be $x=-1.502$, $y=20.0$, 
$\Omega_m=0.9959$ and $\Omega_{\rm rad}=0.004$.
These results are obtained by integrating 
Eqs.~(\ref{beeq1})-(\ref{beeq4})
forward using the lsode stiff integrator. 
It is clear that the matter era is followed by the accelerated epoch with the 
oscillation of $w_{\rm eff}$ around $-1$.}
\label{fig1} 
\end{figure}
\begin{figure}[ht]
\begin{centering}
\includegraphics[width=8.0truecm]{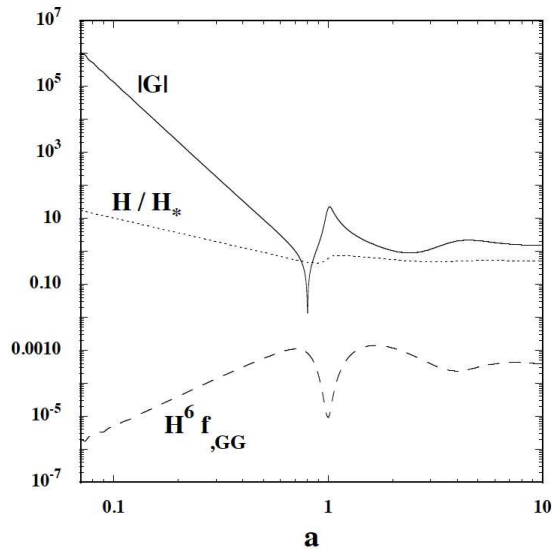} 
 \par
\end{centering}
\caption{The evolution of the quantities $H/H_*$, $|\G|$ and 
$H^6 f_{,\G \G}$ for the same model 
parameters as given in Fig.~\ref{fig1}. 
The GB term changes its sign from negative to positive 
during the transition from the matter era to the accelerated epoch.
The term $H^6 f_{,\G \G}$ becomes much smaller than 1 
as we go back to the past.}
\label{fig2} 
\end{figure}

In Fig.~\ref{fig1} we plot the evolution of $w_{\rm eff}$, $\Omega_{\rm DE}$ 
and $\Omega_m$ for the model (A)
with parameters $\alpha=10.0$ and $\lambda=7.5 \times 10^{-2}$.
The present epoch corresponds to the scale factor $a=1$ with 
$\Omega_{\rm DE}=0.72$ and $\Omega_m=0.28$.
{}From Eq.~(\ref{lamvalue}) there exists a de Sitter point at $y_1=0.518$ 
that satisfies the stability condition (\ref{stacon}).
In fact Fig.~\ref{fig2} shows that the quantity $y=H/H_*$ 
approaches this value in the asymptotic future.

In Fig.~\ref{fig1} the effective equation of state 
$w_{\rm eff}$ oscillates around $-1$ as the system
enters the epoch of cosmic acceleration, 
which implies that the de Sitter solution is a stable spiral.
It is interesting to note that 
$w_{\rm eff}$ drops down to a value less than $-4$
around the present epoch.
We also find in Fig.~\ref{fig2} that the GB term switches its sign 
during the transition from the matter era to the accelerated epoch.
(which corresponds to passing through the minus infinity in logarithmic scale).

\begin{figure}[ht]
 \begin{centering}\includegraphics[width=8.0truecm]{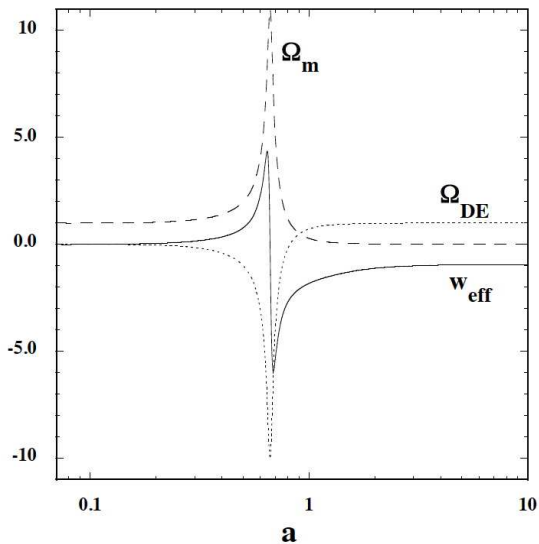} 
 \par\end{centering}
\caption{The evolution of $w_{\rm eff}$, $\Omega_{\rm DE}$ 
and $\Omega_m$ versus the scale factor $a$ for the model (A) 
with parameters $\alpha=0$ (i.e., $f(\G=0)=0$) and $\lambda=1$.
The initial conditions are chosen to be $x=-1.502$, $y=20.0$, 
$\Omega_m=0.9959$ and $\Omega_{\rm rad}=0.004$.
{}From the matter era to the accelerated epoch 
the solution shows an unusual transition where $\Omega_m$
exceeds the order of unity.}
\label{fig3} 
\end{figure}

We have also tried numerical integrations by changing the model
parameters $\alpha$ and $\lambda$.
For the values of $\alpha$ smaller than unity it is not easy to 
to get plausible cosmological evolution.
This is associated with the fact that the stability condition 
(\ref{lamsec}) is difficult to be satisfied around $\G=0$ for smaller 
$\alpha$. In Fig.~\ref{fig3} we illustrate the variation of 
$w_{\rm eff}$, $\Omega_{\rm DE}$ and $\Omega_m$
for the model (A) with parameters $\alpha=0$ and $\lambda=1$.
While a stable de Sitter point exists at $y_1=1.075$, 
$\Omega_m$ becomes larger than the order of unity 
during the transition from the matter era to the 
accelerated epoch.
This reflects the instability of the solutions around 
$\G=0$.
For $\alpha$ smaller than the order of unity, 
the solutions exhibit unusual behavior
similar to that in Fig.~\ref{fig3} or they do not 
reach the de Sitter attractor.

For the model (A) it is difficult to integrate the equations from the redshift 
larger than 50 because the term $1/(96H^6 f_{,\G \G})$ 
in Eq.~(\ref{geper}) gets very large as we go back to the past. 
The decrease of the term $H^6  f_{,\G \G}$ for larger $z$ 
is in fact confirmed in Fig.~\ref{fig2}. 
On one hand, this is a good property, as the spin-2 no-ghost conditions 
are then satisfied \cite{Cal}, in addition to the fact that 
scalar perturbations remain stable \cite{LBM}. 
However, this large mass leads to a rapid oscillation of 
the perturbations $\delta_H$, 
which is hard to be dealt with numerically.
The difficulty of numerical integration is even more severe for the models (B) and (C).
When the equations are integrated forward, we need to choose 
initial conditions carefully so that the oscillating mode $\delta_H^{\rm (osc)}$
is suppressed relative to the matter-induced mode $\delta_H^{\rm (ind)}$.
This property is similar to viable $f(R)$ gravity models in which 
the system is unstable in the high-redshift regime because of 
the appearance of oscillating mode \cite{Star07,Tsuji07}.

\subsection{High-redshift approximate solutions}
\label{subB}

In the regime of high-redshifts one can use an iterative method
(known as the ``fixed-point'' method)
to find cosmological solutions approximately. 
We define $\bar H^2$ and $\bar \G$ to be 
$\bar{H}^2\equiv H^2/H_0^2$ and $\bar{\G} \equiv \G/H_0^4$, 
where the subscript ``0'' represents present values (with $a_0=1$).
The models (A), (B) and (C) can be written in the form
\be
f(\G)=\bar{f}(\G)H_0^2-\bar{\Lambda}\,H_0^2\,,
\ee
where $\bar{\Lambda}=\alpha \lambda \sqrt{G_*}/H_0^2$
and $\bar{f}(\G)$ is a function of $\G$.
The modified Friedmann equation (\ref{be1}) reduces to
\be
\label{Hite}
\bar H^2- \bar H_{\Lambda}^2=\frac13\,
(\bar{f}_{,\bar{\G}} \bar\G-\bar f)-8\frac{{\rm d}
\bar{f}_{,\bar{\G}}}
{{\rm d} N}\,\bar H^4\,,
\ee
where 
\be
\label{Hlambda}
\bar H_\Lambda^2=\frac{\Omega_m^{(0)}}{a^3}+
\frac{\Omega_{\rm rad}^{(0)}}{a^4}+\frac{\bar{\Lambda}}{3}\,.
\ee
Note that $H_{\Lambda}$ represents the Hubble parameter 
in the $\Lambda$CDM model.
In the following we omit the tilde for simplicity.

In Eq.~(\ref{Hite}) there are derivatives of $H$ in terms of $N$
up to the second-order. Then we write Eq.~(\ref{Hite})
in the form
\be
H^2-H^2_\Lambda=
C\!\left( H^2, {H^2}',{H^2}'' \right) \,,
\ee
where $C=(f_{,\G} \G- f)/3-8H^4\, ({\rm d} f_{,\G}/{\rm d} N)$.
At high redshifts ($a \lesssim 0.01$) the models (A), (B) and (C) 
are close to the $\Lambda$CDM model, i.e., $H^2 \simeq H_\Lambda^2$.
We shall introduce an iterative method to derive approximate solutions
in such a regime. 

As a starting guess we set the solution to be $H^2_{(0)}=H^2_\Lambda$. 
The first iteration is then
\be
H^2_{(1)}=H^2_\Lambda+C_{(0)}\,,
\ee
where $C_{(0)}\equiv C\bigl(H^2_{(0)}, {H^2_{(0)}}',{H^2_{(0)}}'' \bigr)$. 
This first iterative solution was used as an approximate solution for inverse curvature gravity
in the paper of Mena {\it et al.}\,\cite{Mena}, while the authors did not 
pursue the idea of iterating the process again.
We shall iterate the process in order to get better approximate solutions. 
The second iteration is
\be
H^2_{(2)}=H^2_\Lambda+C_{(1)}\, ,
\ee
where
$C_{(1)}\equiv C\bigl(H^2_{(1)}, {H^2_{(1)}}',{H^2_{(1)}}''\bigr)$.

If the starting guess was in the basin of a fixed point, $H^2_{(i)}$ will 
converge to the solution of the equation after the $i$-th iteration.
For the convergence we need the following condition
\be
\frac{H^2_{i+1}-H^2_i}{H^2_{i+1}+H^2_i}
<\frac{H^2_{i}-H^2_{i-1}}{H^2_i+H^2_{i-1}}\,,
\ee
which means that each correction decreases for larger $i$. 
The following relation is also required to be satisfied:
\be
\frac{H^2_{i+1}-H^2_\Lambda-C_{i+1}}{H^2_{i+1}-H^2_\Lambda+C_{i+1}}
<\frac{H^2_{i}-H^2_\Lambda-C_i}{H^2_i-H^2_\Lambda+C_i}\,.
\ee
Once the solution begins to converge, one can stop the iteration up 
to the required/available level of precision.

At very high redshifts (say, the epoch of nucleosynthesis), the above method 
is presumably the only one that provides approximate cosmological solutions. 
We have checked that, for $N \gtrsim -4$ (i.e., for the redshift $z \lesssim 50$), 
this iterative method and the direct-forward-integration give the same results. 
To be more precise, the iteration is used in order to find the values 
of $H^2$ and $\G$ at $N=-4$. 
Then these values are adopted as initial conditions for the direct-forward-integrator. 
We integrate the equations for $-4<N<-3$ (where the iterative method still works well) 
and compare $H^2$ as well as $\G$ at $N=-3$ derived by two methods.
We find that these provide identical results with the precision 
of the order of $10^{-9}$.
This shows that the solution derived by direct integration remains close to 
the iterative one at least for all values of $N$ at which the initial guess 
for the iterative method is in the basin of the fixed point.

\begin{figure}[H]
 \begin{centering}\includegraphics[width=8.2truecm]{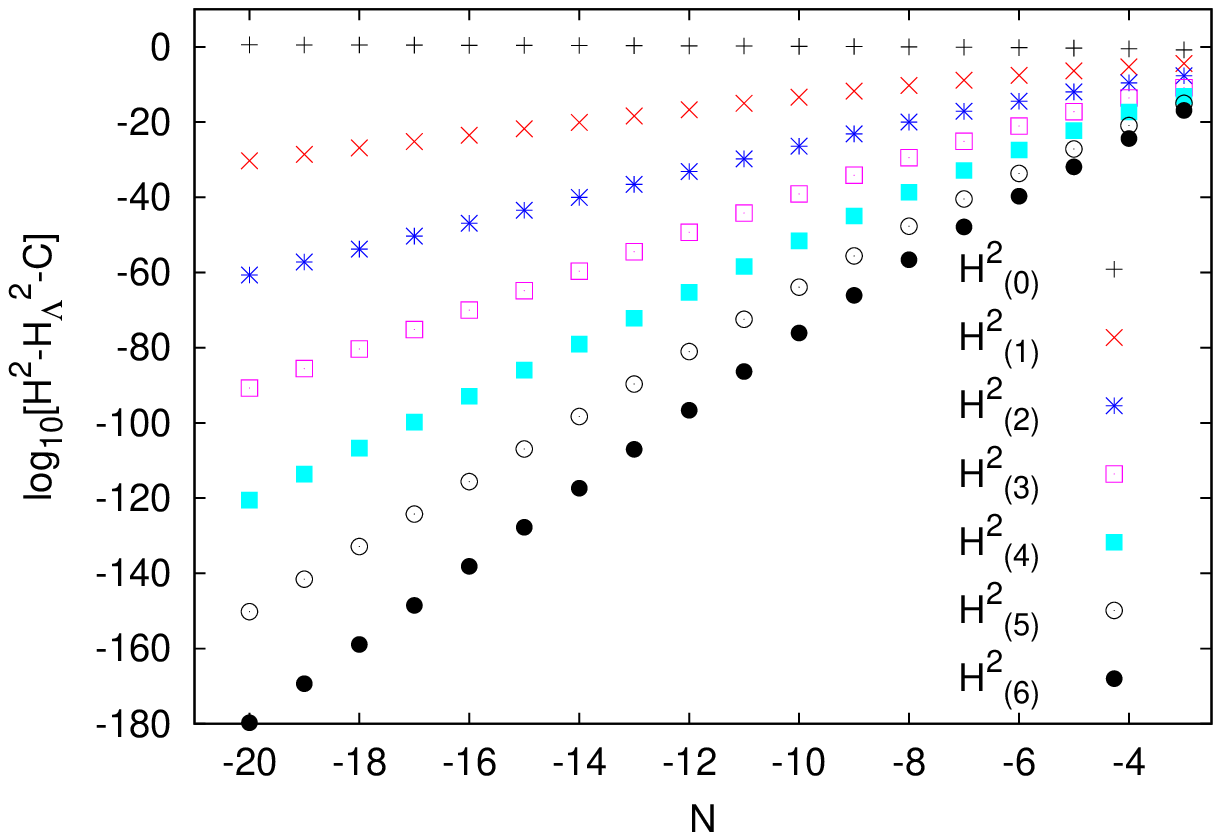} \includegraphics[width=8.2truecm]{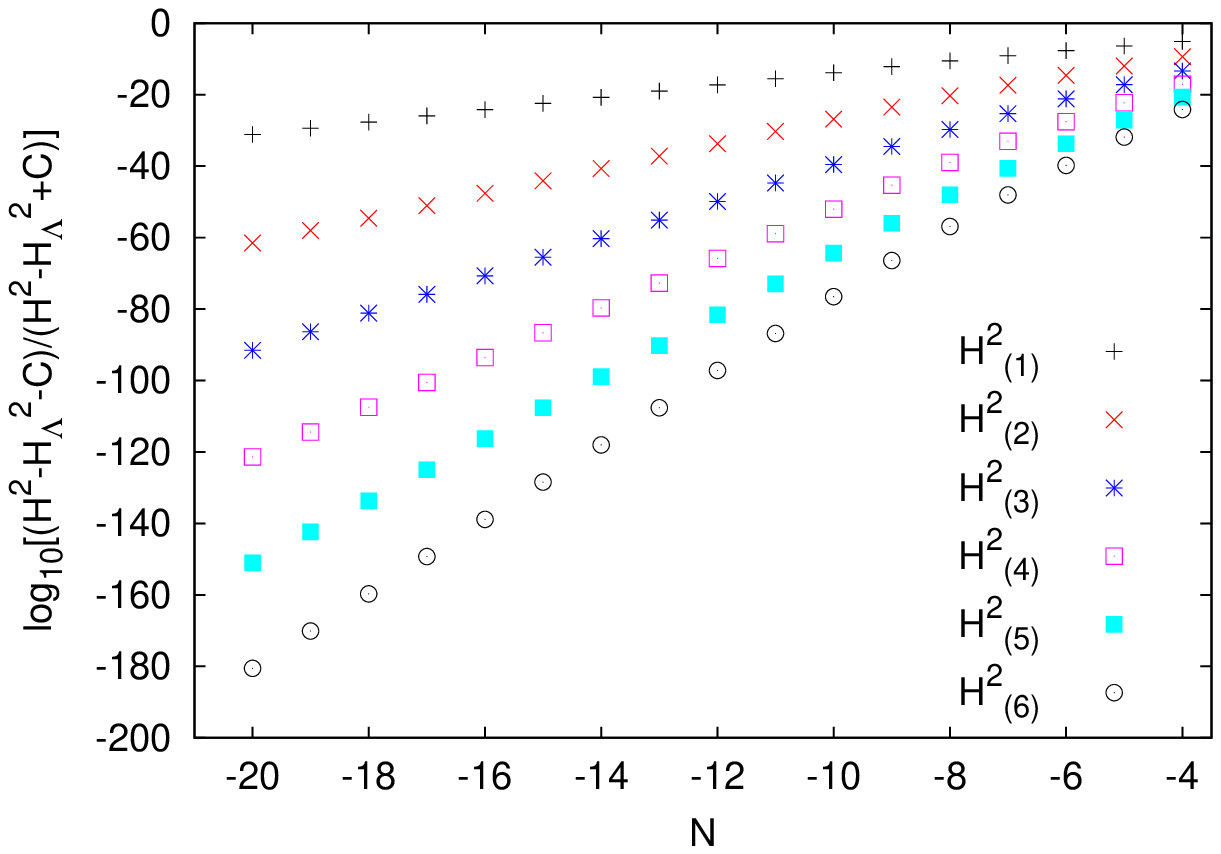} 
 \par\end{centering}
\caption{The plot of the absolute errors $\log_{10}(|H_i^2-H_{\Lambda}^2-C_i|)$ (left)
and $\log_{10}\left[\frac{|H_{i}^2-H_{\Lambda}^2-C_{i}|}{|H_i^2
-H_{\Lambda}^2+C_i|}\right]$ (right)
versus $N$ for the model (A) with $i=0,1,\cdots,6$. 
The model parameters $\alpha$ and $\lambda$ are the same as those in Fig.~\ref{fig1}.
The iterative method provides the solutions with high accuracy in the regime 
$N \lesssim -4$.}
\label{fig4} 
\end{figure}
\begin{figure}[ht]
 \begin{centering}\includegraphics[width=8.2truecm]{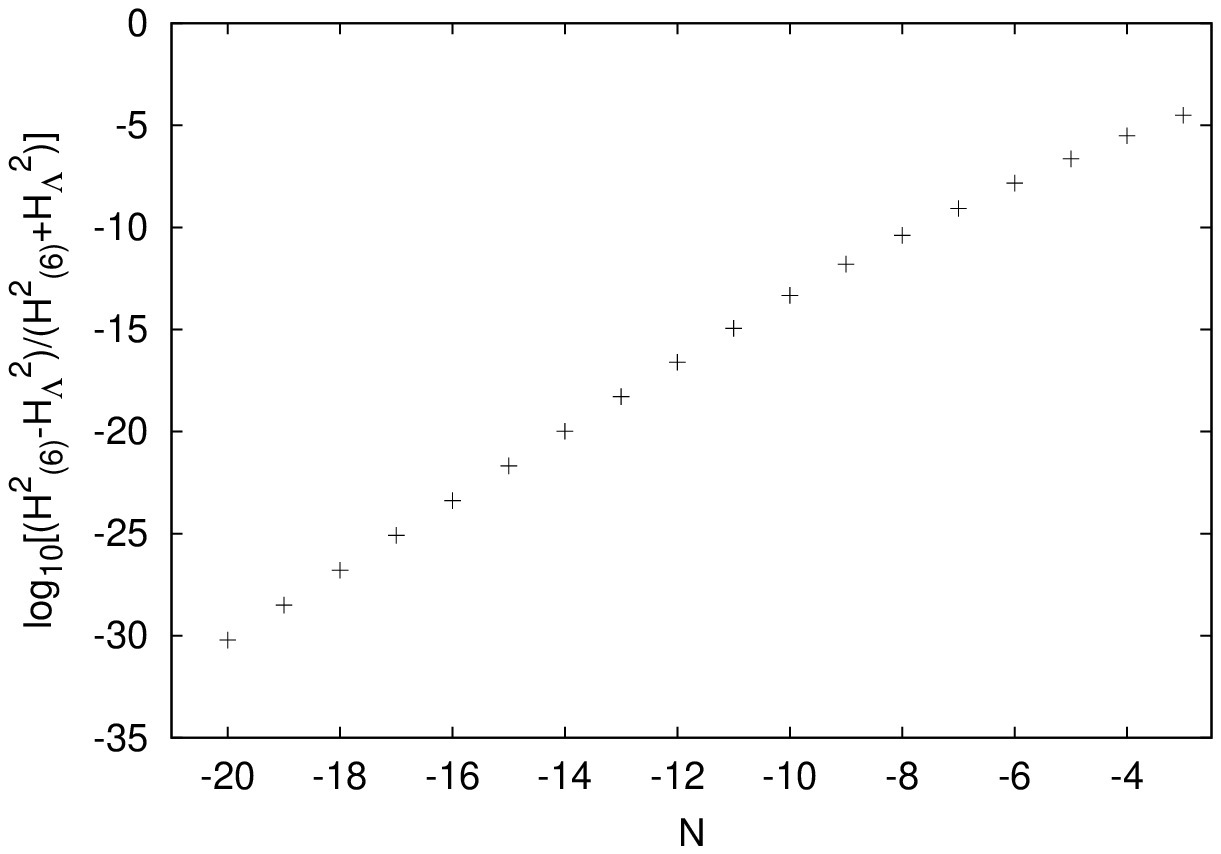} \includegraphics[width=8.2truecm]{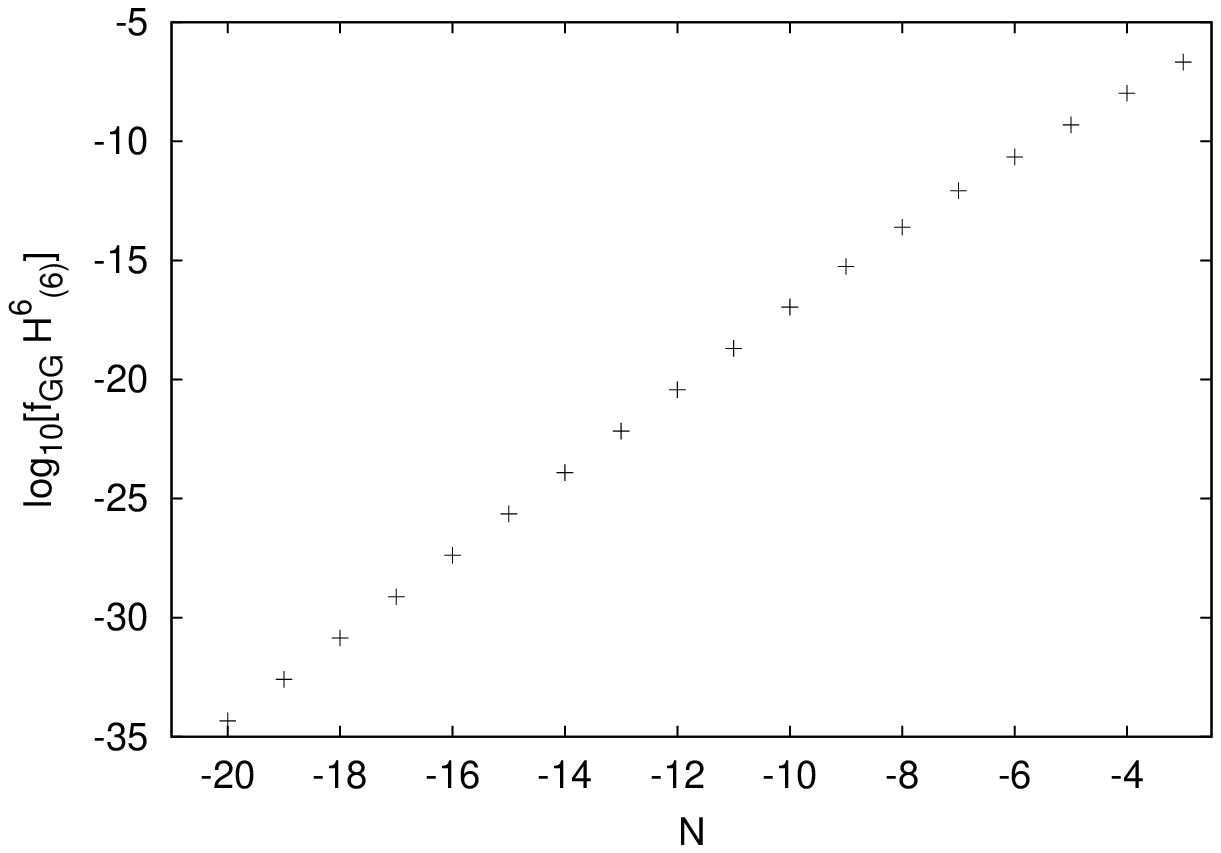} 
 \par\end{centering}
\caption{The plot of the relative error 
$\log_{10}\left[\frac{|H_{(6)}^2-H_{\Lambda}^2|}{|H_{(6)}^2+H_{\Lambda}^2|}\right]$ 
for $i=6$ (left) and the quantity 
$\log_{10} (f_{,\G \G}H^6)$ (right) versus $N$ for the model (A)
with the same model parameters as used in \ref{fig1}.}
\label{fig5} 
\end{figure}

In Fig.~\ref{fig4} we plot the absolute error $\log_{10}(|H_i^2-H_{\Lambda}^2-C_i|)$ as well 
as the relative error 
$\log_{10}\left[\frac{|H_{i}^2-H_{\Lambda}^2-C_{i}|}{|H_i^2-H_{\Lambda}^2+C_i|}\right]$
for the model (A) with $\alpha=10.0$ and $\lambda=7.5 \times 10^{-2}$
(i.e., the same model parameters as used in Fig.~\ref{fig1}).
Note that we have carried out the iteration for 6 times.
The absolute error $\log_{10}(|H_i^2-H_{\Lambda}^2-C_i|)$ is not sufficient to confirm
that $H_i^2-H_{\Lambda}^2$ is really close to $C_i$. However the smallness of the relative error in the left panel of Fig.~\ref{fig5} confirms that the solution derived by the iterative method is very accurate. While this approximation tends to be worse for lower redshifts, the direct integration is well suited for $N \gtrsim -4$ as we presented in the subsection \ref{subA}.

The left panel of Fig.~\ref{fig5} shows that, for $N \lesssim -4$,  
the iterative solution is very similar to the $\Lambda$CDM solution 
characterized by $H^2_\Lambda$.
Hence the Universe passes through the radiation-dominated epoch
to the matter-dominated one as in the $\Lambda$CDM cosmology.
{}From the right panel of Fig.~\ref{fig5} we find that the quantity 
$f_{\G \G}H^6$ is very much smaller than unity for $N \lesssim -4$, 
which leads to an extremely large frequency $M$ 
for the perturbation $\delta_H$.
This is the main reason why we need to use the iterative method
rather than the direct integration to avoid numerical instabilities
in the high-redshift regime.

%%%%%%%%%%%%%%%%%%%%%%%%%%
\section{Conclusions}
\label{conclusions}
%%%%%%%%%%%%%%%%%%%%%%%%%%

In this paper we have constructed viable $f(\G)$ gravity 
models that are cosmologically viable. 
In order to have a stable de Sitter attractor
the condition (\ref{decon}) needs to be satisfied.  For the stability
of radiation and matter fixed points the mass squared $M^2$ given in
Eq.~(\ref{masssqu}) is required to be positive.  These results show
that the quantity $f_{,\G \G}$ must be positive to obtain viable
cosmological evolution.  Since the GB term changes its sign during the
transition from the matter era to the epoch of cosmic acceleration, we
need to construct models in which neither the violation of the
stability conditions nor the divergence of some terms occurs in the
past expansion history of the Universe.

A number of explicit $f(\G)$ models satisfying the above requirements
are given in Eqs.~(\ref{modela})-(\ref{modelc}). 
Even if these models do not require an exotic source of matter responsible 
for the cosmic acceleration, it should be pointed out that in the examples 
presented here a cosmological constant term is still required. 
These models come from the integration of
viable forms of $f_{,\G \G}$ presented in Eq.~(\ref{fGG}).  In the
regime where $|\G|$ is much larger than $\G_*$ (which is the same
order as the present value $\G_0$), the model mimics the $\Lambda$CDM
cosmology.  The deviation from the $\Lambda$CDM cosmology tends to be
important as $|\G|$ approaches the order of $\G_*$.  Since the mass
squared $M^2$ becomes very much larger than $H^2$ as we go back to the
past, this leads to rapid oscillations of the Hubble parameter unless
initial conditions are chosen such that the oscillating mode is
suppressed relative to the matter-induced mode.  The direct
integration of Eqs.~(\ref{beeq1})-(\ref{beeq4}) is prone to numerical
instabilities in the high-redshift regime because of the very heavy
mass $M$.

We have adopted an iterative method to derive approximate solutions in
the high-redshift regime.  The results of Figs~\ref{fig4} and
\ref{fig5} show that the iterative method gives rise to the cosmic
expansion history that is very close to the $\Lambda$CDM model for $z
\gtrsim 50$.  We have used these results as initial conditions at $z
\sim 50$ for the direct forward-integration in the low-redshift
region.  We have found that the effective equation of state $w_{\rm
  eff}$ enters the phantom region ($w_{\rm eff}<-1$) before reaching
the de Sitter attractor with $w_{\rm eff}=-1$ (see Fig.~\ref{fig1}).
For the models with $f(\G=0)=0$ the solutions typically exhibit
unusual behavior where $\Omega_m$ grows larger than 1 during the
transition from the matter era to the accelerated era (see
Fig.~\ref{fig3}).  This is associated with the fact that an
instability around $\G=0$ is present for small $\alpha~(\lesssim 1)$
for the models (\ref{modela})-(\ref{modelc}), while this is not the
case for $\alpha$ larger than the order of unity.

Since the $f(\G)$ models we have proposed mimic the $\Lambda$CDM model
in the high-curvature regime whose energy density is much larger than
the present cosmological one, it should be possible for them to
satisfy local gravity constraints.  We leave detailed analysis for the
compatibility of our models with local gravity experiments for future
work.

%%%%%%%%%%%%%%%%%%%%%
\section*{ACKNOWLEDGEMENTS}
We thank Stephen Davis for useful discussions.
ADF was supported by the Belgian Federal Office
for Science, Technical and Cultural Affairs, under the
Inter-university Attraction Pole grant P6/11. 
ST was supported by JSPS (No.~30318802) and 
by FY 2008 Researcher Exchange Program 
between JSPS and CNRS.
ST is thankful for kind hospitalities during his stays 
in University of London, University of Montpellier, 
APC Paris, IAP Paris and University of Louvain.
%%%%%%%%%%%%%%%%%%%%%

%

\end{document}